\documentclass[]{llncs}
\usepackage{times}  
\usepackage{helvet}  
\usepackage{courier}  
\usepackage{url}  
\usepackage{graphicx}  
\usepackage{float}
\usepackage{amsmath}
\usepackage{csvsimple}
\usepackage{subcaption}
\usepackage{color}

\begin{document}

\title{Studying Migrant Assimilation Through Facebook Interests\thanks{This is a preprint of a short paper at SocInfo 2018. Please cite the SocInfo version.}}

\author{Antoine Dubois\inst{1} \and
Emilio Zagheni\inst{2} \and
Kiran Garimella\inst{3} \and
Ingmar Weber\inst{4}}

\institute{Aalto University \email{antoine-dubois@hotmail.com} \and 
Max Planck Institute for Demographic Research \email{emilioz@uw.edu} \and 
EPFL \email{kiran.garimella@epfl.ch} \and
Qatar Computing Research Institute \email{iweber@hbku.edu.qa}
}


\maketitle

\begin{abstract}
Migrant assimilation is a major challenge for European societies, in part because of the sudden surge of refugees in recent years and in part because of long-term demographic trends.
In this paper, we use Facebook data for advertisers to study the levels of assimilation of Arabic-speaking migrants in Germany,
as seen through the interests they express online. 
Our results indicate a gradient of assimilation  along demographic lines, language spoken and country of origin.  
Given the difficulty to collect timely migration data, in particular for traits related to cultural assimilation, the methods that we develop and the results that we provide open new lines of research  that computational social scientists are well-positioned to address.
\end{abstract}

%
%
\section{Introduction}


Managing migration flows and the integration of migrants is a major challenge for our societies.
Recent crises and conflicts 
have led to large flows of refugees. For example, in 2015 alone, more than one million refugees arrived in Germany, largely from Syria. 

In addition to the challenges of short-term crises, there are also long-term demographic changes that contribute to the migration debate. Longer lives and lower fertility levels  mean that population aging is an inevitable consequence. 
Immigration is often seen as a stopgap measure to address population aging, which would otherwise strain the economy and public finances. In this context, immigration is expected to become a major driver of population dynamics. 

Understanding the processes of assimilation and integration of migrants has become a priority for countries all over the world. Currently, United Nations member states are developing a global compact for safe, regular and orderly migration.\footnote{\url{http://refugeesmigrants.un.org/migration-compact}} This is an inter-governmentally negotiated agreement to cover all dimensions of migration. Policy action is informed by a number of indicators of integration that international organizations produce  \cite{OECDintegration}. 
Traditional indicators include measures of education, 
language acquisition, poverty, 
intermarriage, and other aspects \cite{NAP21746}. These indicators evaluate how 
opportunities and outcomes for migrants and their children
differ from the ones of the host population. 



Cultural assimilation 
is a more elusive quantity to measure 
with traditional methods. 
However, perceived cultural differences 
often fuel negative sentiments towards immigrants. At the same time, immigrants contribute to the cultural diversity of the host society. It is thus important to evaluate processes related to cultural assimilation.

In this paper, we use anonymized, aggregate data from Facebook users, available via Facebook's advertising platform, to evaluate cultural assimilation of Arabic-speaking migrants in Germany. Using this data, we compute an \emph{assimilation score} and compare this score for different migrant populations in Germany, including 
migrants from Austria, Spain, France and Turkey.
To the best of our knowledge, this paper is the first to make use of online advertising data from Facebook to show the feasibility of measuring migrant integration at scale.






%
%
\vspace{-4mm}
\section{Related Work}

Understanding migrant integration and the effectiveness of policy measures to favor assimilation is a longstanding challenge. A wide range of aspects such as `civic integration policies' \cite{doi:10.1080/13691831003764300} or multiculturalism \cite{wright_bloemraad_2012} have been analyzed. 
These studies developed evaluation metrics based on concepts such as political trust \cite{IMRE:IMRE797} as well as  lack of electoral participation or composite measures of civic integration   \cite{wright_bloemraad_2012}. For a review of empirical and theoretical challenges 
see \cite{waters2005assessing}~(2005). 


New information, like Web and social media data, are a main source of innovation in the context of migration studies. Research in this area has focused on using online data to improve estimates of migration flows and stocks. 
After Zagheni and Weber used geo-located Yahoo! e-mail data to estimate international migration flows \cite{DBLP:conf/websci/ZagheniW12}, several platforms have been used to understand the network structure of migration, including Facebook \cite{DBLP:conf/websci/HerdagdelenSAM16} and Google+ \cite{DBLP:conf/asunam/MessiasBWZ16}. Geo-located Twitter data has proved useful for studying the relationship between internal and international migration \cite{zagheni2014inferring}, as well as short-term mobility versus long-term migration \cite{DBLP:conf/websci/FiorioACZWV17,10.1371/journal.pone.0129202}. LinkedIn data has provided insights into global patterns of migration for professionals \cite{statemigration}. 



More recently, Facebook data for advertisers have been used to create estimates of stocks of international migrants \cite{zaghenietal17pdr}. These data 
are a promising source for generating timely migration statistics at different levels of spatial granularity.  
In this paper, we expand the use of these data to study cultural assimilation.

\vspace{-3mm}
\section{Data}

Facebook's advertising platform allows advertisers to programmatically target their ads to a specific population, e.g.\ based on age, gender, country of residence, or spoken language.\footnote{\url{https://developers.facebook.com/docs/marketing-api/targeting-specs/}}
Using the Marketing API, advertisers can obtain estimates of the number of people who belong to a certain demographic group and show certain \textit{interests} (algorithmically generated) based on `likes', pages that they visit and other signals.\footnote{\url{https://www.facebook.com/business/help/150756021661309}}
As an illustrative example, consider \textit{the number of Facebook users speaking Arabic, living in Germany, being in the age group 18--65 and interested in football}. As of May 3, 2018, Facebook Ads manager reports that there are 530k users matching these criteria. 
The complete list of available targeting options is available on the Facebook Ad API page.\footnote{\url{https://developers.facebook.com/docs/marketing-api/buying-api/targeting}}
We used an open source Python implementation to access the Facebook Marketing API and collect such audience estimates.\footnote{\url{https://github.com/maraujo/pySocialWatcher}}

We base our analysis on Facebook's interests.
For each population, we obtain the audience size for each interest and we compare a  measure of prevalence  
across populations. As an oversimplified example, we can compare the fraction of Arabic-speaking migrants in Germany who are interested in typically German interests such as Bundesliga -- the German soccer league -- to the same fraction of Germans in Germany interested in this topic. Arguably, if a migrant population has similar levels of interest as the host population, this is an indicator of assimilation in terms of cultural taste. 

Facebook Marketing API has hundreds of thousands of interests to target ads. In this paper, 
 we collected data for 2,907 interests\footnote{We started with 3,000 interest IDs obtained in the summer of 2017, but 93 of those were subsequently removed by Facebook.} 
with the biggest global Facebook audiences. 
Our main criteria in selecting this subset is that their audiences should be sufficiently large so that the estimates are reliable. 
%
Then, we obtained the audience sizes for those interests for adults (aged 18-65) from different populations (e.g., non-expats living in Germany, or Arabic-speaking expats living in Germany).\footnote{We use the Facebook advertising platform terminology which does not refer to \textit{migrants} but to \textit{expats}, though we use migrant and expat interchangeably.} 
In this paper, we focus mainly on the assimilation of Arabic-speaking migrants in Germany. 
%
To achieve this, we obtained data for (i) Arabic-speaking migrants living in Germany, and (ii) non-expats in Arab League countries.\footnote{A regional league of 22 Arabic-speaking countries \url{https://en.wikipedia.org/wiki/Arab_League}.} 
%
%
Finally, for comparative purposes, we also collected  audience estimates for the same set of interests and demographic groups for other migrant populations in Germany, from Austria, France, Spain and Turkey.

\vspace{-5mm}
\section{Methods: Quantifying assimilation}

Our goal is to obtain an assimilation score that could serve as a proxy for the assimilation of a group of migrants  to a local population in terms of interests expressed by both groups.
To reach this objective, for a given list of interests, we need the audience sizes of three groups: 
(i) the destination country (always Germany in our case), (ii) the target group in the destination country (i.e.\ Arabic-speaking migrants, French migrants, Turkish speakers, etc.), and (iii) the target group's home country (Arab League, France, Turkey, etc). 
We will denote those groups as \textit{Dest}, \textit{Target} and \textit{Home} respectively. 
We will focus on the case where \textit{Dest} is German non-expats.
We then proceed in two steps: (i) we identify `distinctly German' interests -- interests that are more popular in Germany (and more generally in \textit{Dest}) compared to in \textit{Home}; 
 (ii) we use the audience sizes for those interests to compute the assimilation score.
The selection of `distinctly German' interests is a necessary preprocessing step to compute a meaningful  score. If we were showing that migrants had the same level of interests as local people in generic interests that are prevalent across countries, like ``Technology'' or ``Music'', this would not necessarily be a sign of assimilation.

First we describe how, for each of the 2,907 initial interests, we evaluated if they are `distinctly German'. 
Estimating how popular an interest is among a population could be simply done using the percentage of people in that population with the interest. 
However, this approach would be biased as a result of differential online activity level, since more active Facebook users also have more interests. Though the exact methods used by Facebook to calculate these interests is not disclosed, it is known to make use of a user's activity.\footnote{\url{https://www.facebook.com/business/help/182371508761821}} 

To correct for this activity level bias, we can instead use the normalized audience percentage within each population given by
\begin{equation}
\label{eq:interest_ratio}
	IR_p(i) = \frac{A_p\left(i\right)}{\sum_{i=1}^n A_p\left(i\right)},
\end{equation}

\noindent where $A_p(i)$ and $IR_p(i)$ denote, respectively, the audience size and what we call interest ratio for interest $i$ in population $p$. Tables \ref{table:example}(a) and \ref{table:example}(b) show an example of the audiences for three populations and the corresponding interest ratios.
$IR$ can now be compared between populations to obtain typical interests.

\begin{table}[ht]
\begin{subfigure}{0.44\textwidth}
\begin{center}
a)
\begin{tabular}{|l|c|c|c|}
\hline
\textbf{Interests} & \textbf{$A_{Dest}$} & \textbf{$A_{Home}$} & \textbf{$A_{Target}$} \\
\hline
Brewery & 790k & 260k & 14k\\
\hline
Berlin & 6200k & 1500k & 320k\\
\hline
Technology & 1200k & 12,000k& 120k\\
\hline
Music & 1600k & 6400k & 690k\\
\hline
God in Islam & 14k & 21,000k & 170k\\
\hline
\hline
\textbf{Total} & 9804k & 41,160k & 1314k\\
\hline
\end{tabular}
\end{center}
\end{subfigure} 
b)
\begin{subfigure}{0.55\textwidth}
\begin{center}
\begin{tabular}{|l|c|c||c|}
\hline
\textbf{Interests} & \textbf{$IR_{Dest}$} & \textbf{$IR_{Home}$} & \textbf{$IR_{Dest}/IR_{Home}$}\\
\hline
\textcolor{blue}{Brewery} & 0.081 &  0.006 & \textcolor{blue}{13.5}\\
\hline
\textcolor{blue}{Berlin} & 0.632 & 0.036 & \textcolor{blue}{17.6}\\
\hline
Technology & 0.122 & 0.292 & 0.42\\
\hline
\textcolor{magenta}{Music} & 0.163 & 0.156 & \textcolor{magenta}{1.04}\\
\hline
God in Islam & 0.002 & 0.510 & 0.004\\
\hline
\end{tabular}
\end{center}
\end{subfigure} 
\begin{subfigure}{0.95\textwidth}
\begin{center}
c)
\begin{tabular}{|l|c||c|}
\hline
\textbf{Interests} &  \textbf{$IR_{Target}$} & \textbf{$AS_{Target}$} \\
\hline
Brewery & 0.011 & 0.14\\
\hline
Berlin & 0.244 & 0.39\\
\hline
\end{tabular}
\end{center}
\end{subfigure}
\caption{Numerical example of the whole process based on actual data for \textit{Dest} as non-expats in Germany, \textit{Home} as  non-expats in Arab League countries and \textit{Target} as Arabic-speaking migrants in Germany. (a) shows audience size for five interests. (b) shows the normalized interest ratios for those interests for \textit{Dest} and \textit{Home} and the selection of `distinctly German' and `most German' (top 50\%) interests. (c) shows the subset of $IR_{Target}$ interest ratios, as well as the assimilation score for the two selected interests. Typically German interests where $IR_{Target}$ is close to $IR_{Dest}$ correspond to a larger $AS_{Target}$ score.}
\label{table:example}
\vspace{-\baselineskip}
\end{table}

We can identify `distinctly German' interests by comparing the $IR$s for Germany and \textit{Home}. `Distinctly German' interests are defined as those that have a larger $IR$ for Germany than for \textit{Home} (shown in magenta and blue in Table \ref{table:example}(b)). 
For added numerical stability, we extract among those `distinctly German' interests the ones that are the `most German' (shown in blue in Table \ref{table:example}(b)). This is done by dividing the $IR$s for German interests by the ones for \textit{Home}, for each of the `distinctly German' interests, and keeping the top $k$\% interests with highest value of the relative $IR$s.

After selecting the `most German' interests, we compute the $IR$s for the target population only for these interests. 
Then, we define an assimilation score per interest $i$ by dividing the $IR$s for \textit{Target} by the ones of \textit{Dest}. 
\begin{equation}
	AS_{Target}(i) = \frac{IR_{Target}(i)}{IR_{Dest}(i)}
\end{equation}
where $AS_{Target}(i)$ is the assimilation score of \textit{Target} for interest $i$.

For instance, for the interest `Brewery' in the example in Table~\ref{table:example}(c), the assimilation score for the target group is 0.011/0.081=0.14. A specific interest $i$ is considered to be fully assimilated by the migrant population if $AS_{Target}(i) \geq 1.0$. 

Finally, to have a single score for the target population, we aggregate the per-interest scores by taking the median across all the `most German' interests.



%
%
\vspace{-5mm}
\section{Results}
\label{sec:results}


We start by validating some of the assumptions in our data collection.
Although Facebook identifies expats of some countries of origin, such as Spain, most Arab League countries cannot be targeted individually this way, though there is a catch-all ``Expats (all)''. 
Thus we use a proxy for this group by instead obtaining estimates for the Arabic-speaking residents in this ``Expats (all)'' group. 
To test if it is indeed a good approximation for expats from Arab League countries, 
we compared the number (per square km) of migrants in the 16 German states 
from a recent report by the Brookings Institute\footnote{\url{https://www.brookings.edu/research/cities-and-refugees-the-german-experience/}}
with the estimated number of migrants per state using the data from Facebook ads manager. 
We find that there is a near perfect correlation between the two sets of values (Pearson's $r$ = 0.99).
\begin{figure}
\begin{center}
\includegraphics[width=0.65\textwidth]{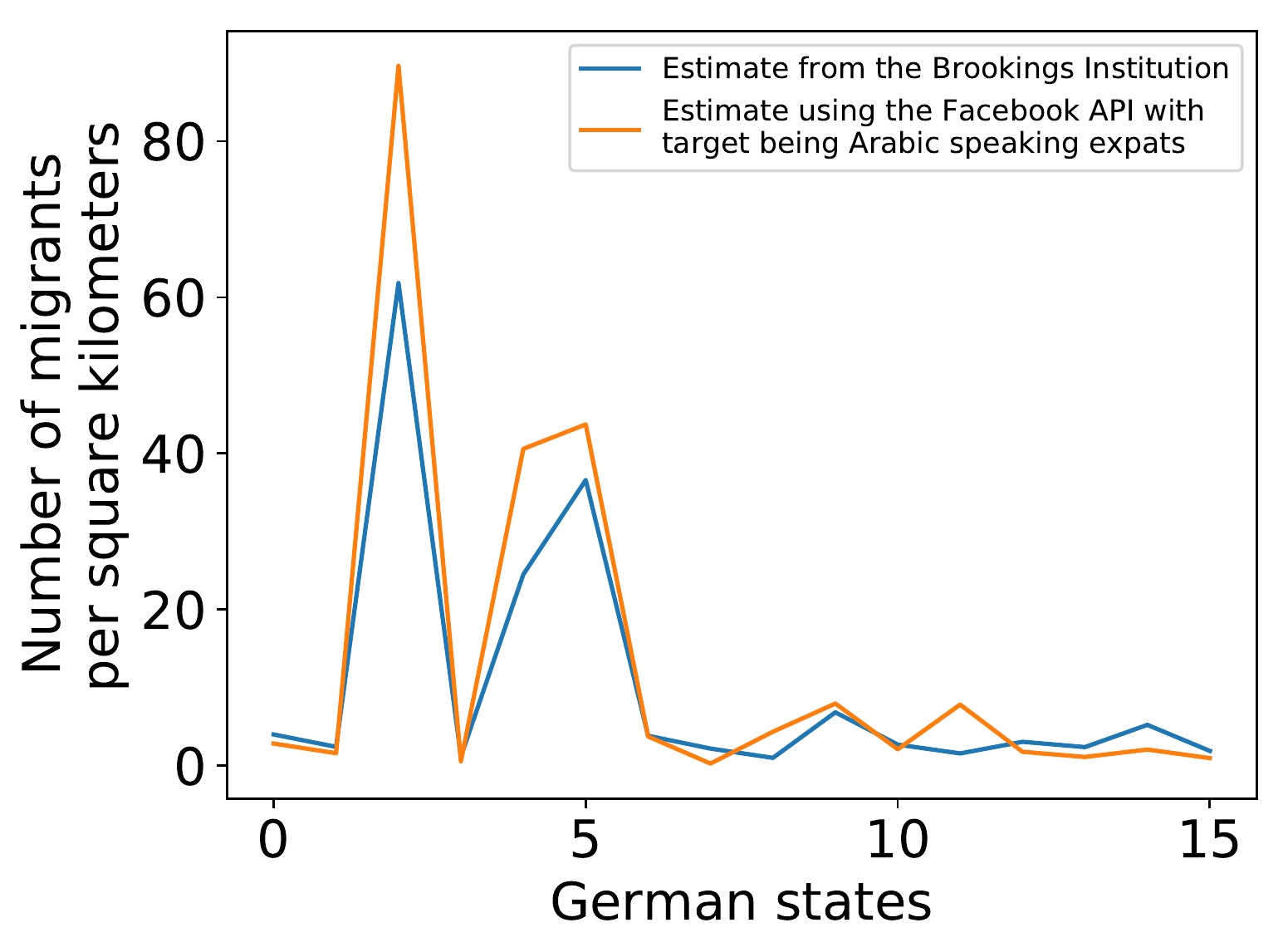}
\end{center}
\caption{The number of migrants per square kilometers in each German state according to two sources; in blue, from a study of the number of refugees by the Brookings institution; in orange, from the estimated audiences by the Facebook ads manager using the Arabic-speaking expats as target population.}
\label{density_comp}
\vspace{-\baselineskip}
\end{figure}

Then, we evaluate the selection of `distinctly German' interests. 
Table~\ref{typical-lists} shows the top and bottom of the list of interests sorted in descending order according to $IR_{Germany}/IR_{\textit{Arab League countries}}$. 
%
%
Note that the top of the list showing some `distinctly German' interests is obtained when using the set of the 2,907 interests that are most popular worldwide.
This is why even more typically German interests such as 
Oktoberfest 
do not show up. 
Similar lists were computed to validate this process when using a different \textit{Home} but are not shown here due to space constraints.

\begin{table}[ht]
\begin{center}
\begin{tabular}{|l|}%
    \hline
    \bfseries Top German Interests
    \begin{filecontents*}{AL_div_bottom_20.csv}
id, Interest
1528,British rock
1743,Die (musician)
2412,Psychedelic pop
2424,Space age pop
2687,Western music (N-A)
2772,Fifty Shades of Grey
2428,Sophisti-pop
2890,Berlin
2020,Premiere
1324,Warner Bros.
1805,Brewery
\end{filecontents*}
\csvreader[head to column names]{AL_div_bottom_20.csv}{}
    {\\\hline\Interest}
    \\\hline
\end{tabular}
\begin{tabular}{|l|}%
    \hline
    \bfseries Bottom German Interests
    \begin{filecontents*}{AL_div_top_20.csv}
id, Interest
1949,CCTV News
2256,Carrefour
2181,Kuwait
2842,Egyptian Arabic
2979,God in Islam
2696,Munhwa Broadcasting Corporation
2910,Insha'Allah
2073,Prophets and messengers in Islam
1738,China Central Television
2878,Arab League
2246,Algeria
\end{filecontents*}
\csvreader[head to column names]{AL_div_top_20.csv}{}
    {\\\hline\Interest}
    \\\hline
\end{tabular}
	\caption{The top and bottom of the list of 2,907 Facebook interests sorted by descending order according to $IR_{Germany}/IR_{\textit{Arab League countries}}$.}
\label{typical-lists}
\end{center}
\vspace{-\baselineskip}
\end{table}

In the remainder of this section, 
we evaluate our assimilation score. We set the `most German' parameter $k$ to 50\% and we recall that this list of interest is always computed using the \textit{Home} corresponding to the \textit{Target} being analyzed.\footnote{We also tested for other values of $k$, 10--50 in intervals of 10, and the trends in the results remain consistent. So we only report results for $k$ = 50.}
%
Our first line of analysis compares \textit{Targets} coming from different countries -- Austria, France, Spain and Turkey.
Since there is a sizable minority of Turkish-speaking non-expats in Germany\footnote{\url{http://bit.ly/2E4UqpD}},
we divide the Turkish population into two: (i) All Turkish speakers in Germany, and (ii) Turkish-speaking non-expats in Germany (this is a subset of (i) containing residents of Germany who speak Turkish).
The assimilation scores for different sub-populations from these countries is shown in the first part of Table~\ref{table:pop_scores} in comparison to Arabic-speaking migrants in Germany. 
The results show that 
European migrants have a higher assimilation score than Arabic-speaking migrants and Turkish speakers. 

\begin{table}[ht]
\begin{center}
\begin{tabular}{|l|c||l|c|}
\hline
\textbf{Target} & \textbf{$AS$} & \textbf{Target} & \textbf{$AS$}\\
\hline
Austrian Migrants & .900  & A: Men & .648\\
\hline
French Migrants & .803 & A: Women & .503\\
\hline
Spanish Migrants & .864 & A: Uni. Grad. & .637\\
\hline
Turkish-Sp. Non-Expats & .922 & A: Not Uni. & .626\\
\hline
Turkish Speakers & .746 &A:  $<$18 & .590\\
\hline
A: Arabic-Sp. Migrants & .643 &A:  18-24 & .665\\
\hline
\hline
A: Men, Uni. Grad., 18-24 & .677 & A: 25-44 & .603\\
\hline
A: Men, Uni. Grad., 25-44 & .620 & A: 45-64 & .504\\
\hline
A: Women, Not Uni., 45-64 & .461 &A:  $>$64 & .553\\
\hline
\end{tabular}
	\caption{Assimilation score ($AS$) for different choice of \textit{Target} in Germany using the top 50\% distinctly German interests of 2,907 interests. Lines 1--5 of the first column correspond to non-Arab populations. The remaining cells, all with the ``A:'' prefix, correspond to Arabic-speaking migrants and sub-groups of this population.}
\label{table:pop_scores}
\vspace{-\baselineskip}
\end{center}
\end{table}

Next, we compared the assimilation scores for different sub-groups among the Arabic-speaking migrants. 
More precisely, we divided this population according to gender, age and education level. 
The results for those sub-groups are shown in the rest of Table \ref{table:pop_scores}.
We explored all attribute combinations to see if there were any additive effects. Due to space constraints, assimilation scores for only a small selection of attribute combinations are included.


Women appear to be less assimilated in terms of their Facebook interests than men. We also observe that the assimilation score for university graduates is slightly higher than that for non-graduates. Note that for both educational levels, the scores are lower than the one for the whole Arabic-speaking migrant group. 
This statistical phenomenon, akin to Simpson's Paradox, is due to the aggregation process of the per-interest scores. Finally, in our analysis, young people between 18-24 are the most assimilated. 
\vspace{-5mm}
\section{Discussion}


The work that we presented in this article has important limitations that we would like to acknowledge. At the same time, we also want to emphasize the potential for further research that computational social scientists can perform in this area.

\begin{figure}
\begin{center}
\includegraphics[width=0.65\textwidth]{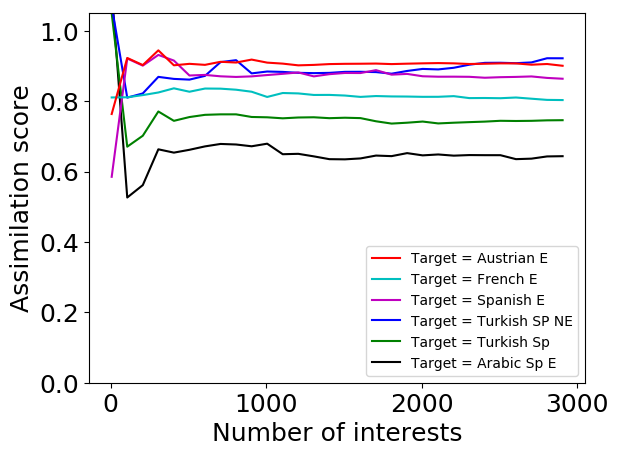}
\end{center}
\caption{Variation of the assimilation score for different Arab and non-Arab populations in Germany using the top 50\% distinctly German interests when changing the size of the set of interests. `E' denotes expats and `NE' non-expats.}
\label{fig:diverse_pops_ev}
\vspace{-\baselineskip}
\end{figure}

Some limitations are related to our methods. For example, the results may be sensitive to the total number of interests that we consider, currently 2,907. 
To check the robustness of our approach, we computed the assimilation scores using random subsets of sizes varying from 100 to 2,900, in steps of 100 within our chosen subset. 
Figure~\ref{fig:diverse_pops_ev} shows the variation in assimilation score. The figure shows that for any subset of size 500 or bigger, the assimilation score becomes stable. The maximum relative change from the average score across all considered populations was 10.4\%, and the average relative change was 4.7\%. 
This indicates that our results are relatively stable with respect to the number of interests being used.

Some limitations are related to the type of data. Our work relies on audience estimates produced by Facebook. The procedures used to infer users' interests are not well documented. For example, there could be differences in how well content in the Arabic or German language is processed, leading to artificial differences in interest profiles. Additionally, Facebook data do not provide information about the number of years that people have spent in the country, a key variable for the study of assimilation processes. 

In our analysis, we grouped together arabic speaking migrants from all 22 countries of the Arab league into a single group, which might introduce biases. Though the arabic speaking countries typically share a lot of cultural similarities, different confounding factors (e.g.\ colonial history) can create differences. Our choice to group these users was because of a limitation by the Facebook Marketing API which does not allow us to target individuals from specific arabic speaking countries (e.g. Syria). 
Also, since a majority (around 75\%) of Arab migrants in Germany are from Iraq and Syria~\cite{deutschland2011bevolkerung}, we think this bias would highly influence our results.

Despite these limitations, this article opens important lines of research that computational social scientists are well-positioned to address with Facebook data for advertisers. First, migration affects the host society and these data can be used to evaluate the extent to which the host society absorbs and embraces exposure to diversity. Second, assimilation processes can be studied at different levels of geographic granularity and in relation to contextual variables like political orientation of various sub-regions. Third, the idea that we presented for the specific case of Germany can be scaled to many countries of the world and used to study macro-regional processes like integration in the European Union. 

\vspace{-3mm}
\section{Conclusion}
We presented a methodology that uses anonymized, aggregate data from Facebook's advertising platform to compare the interest profiles of different migrant groups to that of the German host population. Based on the interest similarities, we derive an \emph{assimiliation score} and observe that this score is lower for Arabic-speaking migrants compared to several European reference groups. We also show that the score varies among sub groups with younger and more educated men scoring highest.
%


%
%

\clearpage 
\bibliographystyle{splncs04}
\bibliography{fb-migrant-integration}{} 
\end{document}